\documentstyle[aas2pp4]{article}

\def\My{\hbox{\kern 0.20em My}}
\def\kms{\hbox{\kern 0.20em km\kern 0.20em s$^{-1}$}}
\def\cmmt{\hbox{\kern 0.20em cm$^{-3}$}}

\received{}
\accepted{}

\slugcomment{to appear in AJ}

\lefthead{Rosado et al.}
\righthead{Sandage 8 and 3 in IC 1613}

\begin{document}

\title{The influence of massive stars in the interstellar medium of IC 1613:  the supernova remnant S8 and the nebula S3 associated with a WO star.}

\author{M. Rosado\altaffilmark{1}, M. Valdez-Guti\'errez\altaffilmark{2},  
L. Georgiev\altaffilmark{1}, L. Arias\altaffilmark{1},  J. Borissova\altaffilmark{3}  and R. Kurtev\altaffilmark{3} }

\altaffiltext{1}{Instituto de Astronom\'\i a,
UNAM, Apartado Postal 70-264, CP 04510,  M\'exico, D. F., M\'exico.}

\altaffiltext{2}{Instituto Nacional de Astrof\'\i sica, Optica y Electr\'onica. Calle Luis Enrique Erro 1, 72840 Tonatzintla, Pue.,
      M\'exico. }
\altaffiltext{3}{Institute of Astronomy, Bulgarian Academy of Sciences. 72
      Tsarigradsko chauss\`ee, BG --1784 Sofia, Bulgaria. }

\begin{abstract}

We present a detailed kinematical analysis of two selected nebulae in the 
Local Group irregular galaxy IC 1613. The nebulae are: S8 (Sandage 1971), the 
only known supernova remnant in this galaxy, and S3, a Wolf-Rayet nebula associated 
with the only WO star in this  galaxy. 
For S8, 
 we have obtained and analyzed its radial velocity field, where we
found complex profiles which can be fitted by several velocity components.
These profiles also show the presence of high velocity, low density gas. From this, we have obtained the expansion velocity, estimated the preshock density and calculated the basic kinematical parameters of this SNR.
We suggest that in S8 we are seing a SNR partially hidden by dust. This suggestion
comes from the fact that the SNR is located between two superbubbles where a
ridge of obscured material unveils the existence of dust. Moreover, we show 
that this hypothesis prevails when energetic arguments are taken into 
account. In the case of S3, this nebula shows bipolar structure. By means of its kinematics, we have  analyzed its 
two lobes, the ``waist'', as well as its relation with the nearest 
superbubbles.
For the first time we are able to see closed the NW lobe, showing a clover 
leaf shape. This fact allows a better quantitative knowledge 
of the nebula as a whole. Furthermore, we found evidence of an expansion 
motion in the NW lobe.
In the light of our results, we can express that these nebulae are 
the product of very massive stellar evolution. It is surprising the 
influence these stars still have in shaping  their surrounding
gas, and on the energy liberation towards the interstellar medium of this galaxy.
 
\end{abstract}



\section{Introduction}

The irregular galaxy IC 1613 is a faint galaxy of the Local Group located at 
a distance of 725 kpc  (Freedman 1988a, b). Because of its proximity, it is an
exceptional target for the study of the interrelationship between gas and 
stars within this galaxy. 

In two previous studies on this galaxy (Valdez-Guti\'errez
et al. 2000 (hereafter Paper II) and Georgiev et al. 1999 (hereafter Paper I)) we have discovered, by means of  H$\alpha$
and $[SII]$ Fabry-Perot interferometry,  that the ionized gas is distributed 
in large diameter, expanding,  ring-shaped structures (superbubbles) which cover
the whole optical dimension of the galaxy. 
Moreover, the NE region of this galaxy contains the brightest superbubbles and it is
assumed that very recent star formation is concentrated in this region rich 
in gas. The superbubbles found outside this region are dimmer, indicating 
that the gas is much less dense there. 

As studied in Paper I and in Hodge (1978)
by means of color-magnitude diagrams of the stellar associations, we have also shown that the superbubbles are
physically linked to massive star associations and that they are systematically older than the
dynamical ages of the superbubbles. This fact, and some energy considerations, are indicative that the 
detected superbubbles are formed by the combined action of winds and 
supernovae explosions of the more massive stars of the interior associations.
However, the supernova explosions traces cannot be detected in the classical ways, because they are too old according to the dynamical timescales
of the superbubbles.
 
These previous studies allow us to present an overview on the relationship
between stars and gas in IC 1613. However, we thought it worth to study in detail
some of the nebulae of this galaxy, taking into account their environment. 
There are two nebulae of IC 1613 that revealed to be very interesting: the only supernova remnant  
detected in this galaxy, the supernova remnant Sandage 8, S8, located at the NE
region of IC 1613, and the nebula hosting the only Wolf-Rayet star detected in
this galaxy, Sandage 3 (Sandage, 1971), S3, located at the south and outside 
the NE quadrant. In this work we study the kinematics of the ionized gas in
these nebulae.

\section{Observations and data reduction}

The observations were carried out with the UNAM Scanning Fabry-Perot (FP)
interferometer  PUMA (Rosado et al. 1995) on December 5-6 1996 and November 15 1998.
This instrument is currently in use at the f/7.9 Ritchey-Chretien focus 
of the 2.1 m telescope at the Observatorio Astron\'omico Nacional at San 
Pedro M\'artir, B.C., M\'exico. These observations are the same that we 
used in our study of the HII region and superbubble kinematics 
(Paper II). 

In the same way, the photometric calibration of the nebulae was carried out
as described in Paper II where
 H$\alpha$ surface brightnesses and fluxes of the selected nebulae are listed in Table 4 of that
work. The fluxes amount to 2.88 $\times10^{-13}$  and 7.24 $\times10^{-13}$ 
\mbox{ergs cm$^{-2}$ s$^{-1}$} for S8 and S3, respectively, with an error of about 20 $\%$.

\section{The Supernova Remnant S8 in IC 1613.}

This supernova remnant (SNR) is the only object of this class known in IC 1613.
It is located in the complex of bright superbubbles in the NE region of
this galaxy and studied in Paper II. 
Indeed, it is seen projected within superbubbles R9 and R10 (at the boundary of
this latter), following the notation of Paper II, and 
in the vicinity of R4. It appears catalogued by Sandage (1971) as S8, by Hodge et al. (1990) as the HII region number 49 and in Paper II as N18. Its nature as SNR has been
established on the following grounds. Indeed, its optical emission shows high [SII]/H$\alpha$ line-ratios (d'Odorico,
Dopita \& Benvenuti 1980, Peimbert, Bohigas \& Torres-Peimbert 1988), it has a
non-thermal radio spectrum (Dickel, d'Odorico \& Silverman 1985), it is an
extended source of X-ray emission and it shows important velocity gradients and
internal motions in its radial velocity field derived from the H$\alpha$ line
(Lozinskaya et al. 1998). 
This latter work, corresponding to a remarkable and complete multi-wavelength
study of S8, also gives the internal motions of this SNR revealing the 
existence of several velocity components in the SNR's radial velocity profiles.
According to Lozinskaya et al. (1998), the main component appears to be 
quite broad ($FWHM = 270 \pm 3$ km s$^{-1}$) and shows a systematic velocity 
gradient from --290 km s$^{-1}$ to the east to --340 km s$^{-1}$ to the west of 
S8 (heliocentric velocities). 

 However, from the analysis of the radial
velocity field and/or broadenings of the several velocity components, these authors do not find any
direct indication of an expanding shell motion. 
Indeed, there is no indication of a variation of the difference in velocities 
of the components nor in the broadenings of the velocity components from the 
center to the boundaries, characteristic of radial expanding shells. Given that
S8 is one of the most conspicuous nebulae in our kinematic observations of 
IC 1613 and that our observations have better spectral resolution (35 versus 
114 km s$^{-1}$) and similar spatial resolution as those of Lozinskaya et al.
(1998), we undertook the study of the detailed radial velocity field of S8 with
the aim to find some direct evidence of expansion from our kinematic data.

Figures 1 and 2 show the location of S8 relative to the bright network of 
expanding superbubbles studied in Paper II and also, the location of the Wolf-Rayet nebula S3 that we will discuss in Section 4. Figure 1 
shows the field at H$\alpha$ while Figure 2 at [SII]. The images correspond to 
 the FP velocity maps in H$\alpha$ and [SII] at the 
heliocentric velocity of --253 km s$^{-1}$.
Two bright and small HII regions, Sandage (1971) regions S7 and S6,  are located near the SNR to the north and 
the NW, respectively. 
The SNR S8 and the HII regions S7 and S6 appear located at the boundary of 
the superbubble R10 (catalogued in Paper II). 
It is interesting to note also that between the superbubbles R4 from one side 
and R9 and R10 from the other side,  a ridge of obscured material is detected
suggesting the presence of dust.

Figure 3 shows a close-up of the SNR emission at H$\alpha$. The S8 region is seen as having dimensions of 
8$\arcsec$ $\times$ 6$\arcsec$ (corresponding to a mean linear radius of about 12 pc at a 
distance of 725 kpc). It is possible that due to the low angular resolution, no filamentary structure is detected. The brightness distribution has a peak toward the center instead of limb-brightening as it is expected for shell-type SNRs. On the other hand, the spectral index of the nonthermal radio emission is $\alpha$ = -- 0.6, typical of shell-type SNRs. Therefore, it is possible that S8 is only the brightest part of a larger
 diameter shell-type SNR that extends further to the
NE, quite near the intersection of superbubbles R4 and R9, where a dust ridge 
is suggested. 

The presence of dust there can dim the possible optical and X-ray emission 
below the detection limits. It is interesting to note that in the VLA 
observations of Lozinskaya et al. (1998) the SNR is located at the northern 
edge of their field and consequently, these radio observations did not cover 
the NE field where the hypothetical other part of the shell could be located. 
A nearer equivalent to the SNR S8 would be the SNR N63A in the LMC 
(Chu et al. 1999). N63A is seen projected within the boundaries of the 
superbubble N63 of larger dimensions. In the case of N63A, the bright optical 
emission only occupies about  the northwestern quarter of the true SNR extent
revealed by X-rays and radio-continuum emissions.  However, better spatial 
resolution optical observations with the HST (Chu et al. 1999) revealed the 
existence of faint and diffuse cloudlets located all over the boundaries of 
the SNR. Some of these cloudlets show radial density gradients suggesting 
cloud evaporation. Thus, it is possible that the optical emission of S8 is 
also due to evaporating cloudlets shocked by the primary blast wave of the SNR
that produces the X-ray emission. 
It is also possible that S8 is the remnant of a supernova (SN) explosion inside a superbubble (in this case, the 
superbubble R9) as Lozinskaya and collaborators pointed out. This fact could 
also explain the unusual shape of the SNR's emissions. However, how could a 
dense fragment remain inside the superbubble?. Why we do not see the interaction with the dense shell?.

Our kinematic data in the lines of H$\alpha$ and [SII]($\lambda$ = 6731 \AA) 
allows to obtain the radial velocity field of this SNR. We 
extracted radial velocity profiles for each pixel (equivalent to 1$\arcsec$.19 for  
H$\alpha$ observations and to 2$\arcsec$.38 for [SII] observations) of the SNR and 
neighboring regions. Typical SNR radial velocity profiles, in  H$\alpha$ and 
[SII], are shown in Figure 4. These profiles are complex and, consequently, imply the existence of 
several components at different velocities. We made a profile fitting of two 
or more Gaussian functions as already described in Paper II. In general, the  accuracy in the peak velocities of single velocity profiles is better than $ \pm 4$ km s$^{-1}$ while the error in the FWHMs of about $ \pm 3$ km s$^{-1}$. In this case, the accuracy in the determination of the peak velocities of the several velocity components and the error in the FWHMs are not as good as in the case of the single profiles because of the complexity of the velocity profiles and the broadening of the different velocity components. We have estimated the errors in the determinations of  the peak velocities and FWHMs  by averaging the obtained values of these quantities in several alternative profile decompositions of the same velocity profile. In this way we obtain that, for the more complex profiles, the errors in the determination of the peak velocities of the different velocity components are of about $ \pm 12$ km s$^{-1}$ while the errors in the FWHMs are of about $ \pm 15$ km s$^{-1}$.

The radial velocity field of S8, obtained from our FP 
observations, is shown in
Figure 5. In this figure we have marked the main components, product of our profile fitting. We have also marked the boxes over which the velocity profiles were integrated. In addition to the main components, we also detect
high velocity gas, red and blueshifted relative to the HII region velocity. 
As Lozinskaya et al. (1998) have already  found, the intensity of these 
components is about 10 to 20 percent of the intensity of the main 
components. Also, for the high velocity gas, there is no  clear expansion 
pattern.

The H$\alpha$ velocity profiles show the presence of gas at the heliocentric 
velocities between --234 $ \pm 12$ km s$^{-1}$ and --253  $ \pm 12$ km s$^{-1}$. The FWHM of this component is of about 66  $ \pm 10$ km s$^{-1}$. This gas reaches almost the heliocentric 
velocity of the HII gas of the galaxy that, in the neighborhood has 
velocities between --235  $ \pm 4$  and --242  $ \pm 4$  km s$^{-1}$. 
The intensity of this component varies between 4.98 $\times$ 10$^{-5}$  to  1.67 $\times$ 10$^{-4}$  ergs cm$^{-2}$ s$^{-1}$ sr$^{-1}$. 
A second component, much broader, has velocities between --317  $ \pm 12$ and --342  $ \pm 12$ 
km s$^{-1}$. The FWHM of this component varies from 66  $ \pm 15$ to 132  $ \pm 15$ km s$^{-1}$. 
The intensity of this component varies between 2.51 $\times$ 10$^{-5}$  to  2.1 $\times$ 10$^{-4}$  ergs cm$^{-2}$ s$^{-1}$ sr$^{-1}$. 
 
This component corresponds to Lozinskaya's et al. (1998) main component but we 
find that the FWHM is about half of the value reported by those authors well above the error of  $ \pm 15$ km s$^{-1}$ estimated for this quantity.
In addition,  we do not detect a gradient in the heliocentric velocity of
the main component (at about --300  km s$^{-1}$) across the SNR's minor axis as 
Lozinskaya et al.  (1998) do. On the other hand, the broadening of this 
component is higher at the photometric center of S8 and to the north and west 
of the center but no clear expansion pattern is appreciated. The [SII] 
velocity profiles show the same features described above.

All these data, collected by us and by the authors cited before, suggest 
that S8 could be formed by the interaction of a primary blast wave with a 
dense clump of large dimensions. The optical emission of S8 is probably due 
to the radiative shock induced in the clump according to McKee \& Cowie 
(1975) scenario. 
In this context, the broadening of the velocity profiles is 
a measure of the velocity of the shock induced in the dense cloud. A typical 
value of the full width at zero intensity (FWZI) of the main components of the velocity profiles 
 (i.e., without taking into account the components at high velocities) is of 
about 380 km s$^{-1}$ implying induced shock velocities, V$_{c}$,
of 170 km s$^{-1}$ for the dense material where the main velocity components 
discussed above are emitted. This value is in agreement with the velocity 
values obtained by Lozinskaya and collaborators for the bright emitting gas 
and with the value derived by Peimbert et al. (1988) 
from the emission line spectrum. 
The X-ray emission is due to the heating of the less dense gas where the 
primary blast wave is propagating at larger shock velocities as described 
by Lozinskaya et al. (1998). An estimate of the preshock density, n$_{0}$, 
could be derived from the electron density obtained from the 
observed [SII] 6717/6731 line-ratio,  n$_{e}$[SII] -- an estimate of this quantity from the observed H$\alpha$ flux of the SNR is not reliable because of the poor angular resolution of all ground based observations carried out till now. 

Thus, in order to estimate the preshock density we use the fit to the Raymond's (1979) radiative shock models:

 n$_{e}$[SII] = 31.48   n$_{0}$ ( V$_{s}$/100 km s$^{-1}$)$^{2}$
   
(Cant\'o, private comm.) where  n$_{e}$[SII] and  n$_{0}$ are expressed in units of cm$^{-3}$ and  V$_{s}$, the shock velocity, is expressed in units of km s$^{-1}$. 
Using our derived value for  V$_{c}$ as  V$_{s}$ and taking n$_{e}$[SII] = 1300 - 1500 cm$^{-3}$ (corresponding to 
the values of the electron density obtained by Lozinskaya et al. 1998 and 
Peimbert et al. 1988, respectively), we obtain  
n$_{0}$ = 14 - 16 cm$^{-3}$. This estimate for the preshock density should be considered as an upper limit because it is based only on the spectroscopy of the brighter (and denser) cloudlet of S8.

We can also obtain an upper limit to the  energy released by the SN into the interstellar medium, E$_{0}$, by
substitution of the adopted values: R$_{s}$ = 12 pc, V$_{s}$ = 170 km s$^{-1}$ 
and n$_{0}$ = 15 cm$^{-3}$ in the Sedov's relation  
E$_{0}$ = 1.37 10$^{42}$ n$_{0}$  $V_{s}^{2}$ R$_{s}^{3}$ (Cox 1972), where E$_{0}$ is in ergs,  n$_{0}$ is in cm$^{-3}$, V$_{s}$ is in  km s$^{-1}$ and R$_{s}$ is in pc. 
We obtain  E$_{0}$ = 1.0 10$^{51}$ ergs. This estimate fall 
within the typical ranges of  E$_{0}$ in supernova explosions (from 
10$^{50}$ to a few 10$^{51}$ ergs).

If S8 were the brightest part of a larger SNR  which extends further to the NE till the obscured rim at the boundary of R9, the average preshock density should be lower than the assumed one while the radius should be 3 times larger. It would  still be possible to obtain  energy values typical of SN explosions.
The age, if this latter assumption were right, would be 3 times longer than the
value of (3 - 6) 10$^{3}$ yr estimated by Lozinskaya et al. (1998).

A better understanding of this SNR requires: a) radio continuum observations of
high sensitivity in a field of at least 30$\arcsec$ centered at the S8 position in 
order to try to detect a complete shell of non-thermal radio emission, 
b) better angular resolution images and spectra at optical wavelengths, 
ideally with the HST.

In conclusion, S8 is a SNR evolving in a clumpy medium where active star 
formation and stellar winds of massive stars form large superbubbles. 
Its strange morphology in different wavelengths could be due either to the 
interaction with a pre-existing superbubble, as Lozinskaya et al. (1998) have 
suggested, or to the fact that only the brightest region of a larger SNR shell is 
detected, as suggested in this work. 
In either case, the primary blast wave, moving at a velocity of thousands of
km s$^{-1}$ in a rarefied medium, induces radiative, secondary shocks in more 
dense clumps that produce the optical emission.

\section{ The Wolf Rayet star embedded in the HII region S3.}

This  Wolf-Rayet (WR) star was discovered by d'Odorico \& Rosa (1982) and later on
it appeared in the list  of eight stars in IC 1613 that  Armandroff \& Massey (1985) considered as candidates to WR stars. It was the number 6 in this list (WR6). Subsequent studies (Azzopardi et al. 1988) have shown that only WR6 is indeed a WR star, while the other candidates were classified as massive supergiants of spectral 
types ranging from OB to A.
Thus, WR6 (according to the nomenclature of  Armandroff \& Massey 1985) 
seems to be the only known WR star in IC 1613. 
This WR is located in the HII region S3 in Sandage (1971), named V37 in Hodge
(1990) and N9 in Paper II. This star has also been 
 studied by Davidson \& Kinman (1982), Massey et al. (1987), Lequeux et al. (1987)  and Kingsburgh \& Barlow (1995) while the nebula associated with this WR star has been studied by Goss \& Lozinskaya (1995) and Afanasiev et al. (2000).

Kingsburgh \& Barlow (1995) classified this star as an oxygen WR star, WO. 
There are only 5 stars known in the Local Group belonging to this class and 
WR6 is the brightest member of this class. This class of WR stars constitute a
sequence different from the WN and WC sequences, and represents the late
 stages of very massive (M$_{*}$ $>$ 40 M$_{\odot}$) stellar evolution where He and C
burning is taking place at the stellar core. 
Taking into account that this WO star is inside the nebula S3 and, that it 
could be the most important source of ionization of two of the  superbubbles catalogued 
in Paper II: R16 and R17, we undertook the detailed 
study of the interstellar medium (ISM) in the vicinity of this star, mainly of
its kinematics, with the aim of knowing more about the interaction of this 
massive star with its surrounding ISM.

In our field, S3 - the HII region associated with WR6 - is well resolved and
located at the southern part of the NE region of IC 1613 at the position 
(1950): 1 02 27.3 and +01 48 17 (Armandroff \& Massey 1985).  Its observed 
equivalent diameter is 253.6 pc (see the definition in Paper II).
According to Hodge et al. (1990), S3 is composed of 5 HII regions (a-f).

S3 is located near the southern border of our 10$\arcmin$ field as it is shown in Figures 1 and 2. 
Figure 6 shows a close-up of the monochromatic H$\alpha$ image shown in Paper II.
In this figure, the morphology of S3 looks spectacular; it seems to be 
interacting with at least a couple of superbubbles: R17 that is located 
at its NW side and R16 farther away.
On the other hand, its shape recalls a large bipolar nebula, 
having the WO star at its center, like the galactic planetary nebula NGC 2346 (Arias et al. 2000).  
In this context, the nebula S3 itself would be the central part of a bipolar 
structure whose NW lobe encompasses the superbubbles R17 and R16.

In Paper II we have discussed  that the superbubbles R16 and 
R17 are probably linked to  the stellar associations H8 and H9 (Hodge
1978). By applying the same automatic method described in Paper I, the new boundaries of the stellar associations are outlined (Borissova et al. 2001). As in the case of the NE region (Paper I), the Hodge (1978) associations H8 and H9 divide into several smaller groups. The `new' associations in the region are located almost coincident with the superbubbles. As illustrated in Figure 6, the associations are located within or at the peripheries of the superbubbles' boundaries. This suggests that the winds of the massive stars of the interior associations have probably formed these superbubbles and that the associations found at the periphery were formed by the self-induced star formation mechanism.
 However, the ionizing flux of the close WO star could also contribute to the ionization of these superbubbles. It is
interesting to note that the WO star does not belong to any of the stellar associations currently identified.

Figure 7 and Figure 8 present the H$\alpha$ velocity maps where the S3 complex shows some emission. Both
figures are derived from the same data, but the velocity maps shown in Figure 8
 were smoothed both spectrally and  spatially and the contrast has been lowered in order
 to enhance the visibility of the faint filaments located at the NW of S3. The
superbubbles R16 and R17 of Paper II (whose boundaries are marked in Figure 6) come to be noticed in some of the velocity maps of Figure 8.
As one can see in Figure 7, S3 is composed of two lobes (NW and SE) sorting out from
 a bright waist which appears like a continuation of the NW lobe. In fact,
it appears that the waist, that hosts the WO star in its brightest region, is not a
 torus (unless it is seen perpendicular to the line of sight). The SE lobe shows
 more filaments and knots than the NW one and its dimensions are about half of
 the dimensions of the NW lobe. The NW lobe ends in its brightest region, in the
waist. In its brightest region, it appears as an incomplete ellipse, but we
 cannot see from these velocity maps if it is closed or open. The NW lobe
 appears crossed by the superbubble R17 while the superbubble R16 is located
outside of it. The waist has a small bubble-like feature (hereafter called Bubble A) at its SW end of dimensions: 13$\arcsec$.2 x 9$\arcsec$.2. In
 Figure 8 the details of the internal structure of S3 are not seen because
we favored to show the quite faint external superbubbles.

Figure 9 shows
the [SII] ($\lambda$ 6731 \AA) velocity maps of the same region. In this
figure, there is no evidence of the superbubbles R17 and R16. Most
 of the [SII] emission corresponds to the bipolar structure internal to S3. For the first time,
we detect the NW lobe closed (see the velocity map at V$_{HEL}$ = --233 km s$^{-1}$). This fact allows us to have a better knowledge of the
 dimensions and shape (not exactly elliptical but in the form of a clover leaf) of the NW lobe. The SE
 lobe is as faint as the NW lobe. The Bubble A,
at the end of the waist, is also detected. The dimensions of the different
 regions are: 60$\arcsec$ $\times$ 50$\arcsec$ or 214 pc  $\times$ 179 pc for the NW lobe, 34$\arcsec$  $\times$ 16$\arcsec$ or 122 pc  $\times$ 60 pc for the SE lobe and 22$\arcsec$  $\times$ 12$\arcsec$ or 81 pc  $\times$ 45 pc for the dimensions of the waist. In all these determinations we have adopted the distance to IC 1613 as 725 kpc, as mentioned in the Introduction. The NW lobe is better appreciated in the velocity maps
 corresponding to V$_{HEL}$ from --233 to --213 km s$^{-1}$ implying that the systemic
 velocity of the bipolar structure must fall within these velocity values.

In order to study the kinematics of the lobes we have obtained H$\alpha$ and
 [SII] radial velocity profiles integrated over boxes of 10  $\times$ 10 and 5  $\times$5
 pixels, respectively, covering the S3 complex.   Scanning Fabry-Perot observations allow to identify the possible contamination of the radial velocity profiles by possible emission line stars seen projected in the same direction than the integration boxes. We have analyzed this possibility; such a profile could be identified easily because, in that case, the continuum emission of the star would also affect the level of the continuum of the profile. In addition, the effect of an emitting line star would affect only one of the points of the obtained 2-D velocity field and, consequently, it is also possible to identify any abrupt change in the velocities of the velocity components due to this possibility. Except for one velocity profile where the continuum level was higher, we did not find that this effect was relevant in our velocity determinations.

The  H$\alpha$ velocity profiles of the brightest regions are, in general,
 simple and can be fitted by a single Gaussian function of FWHM, varying between
 66 and 81 km s$^{-1}$ and peak velocities ranging from V$_{HEL}$= --223  $ \pm 8$ km s$^{-1}$ to --243  $ \pm 8$  km s$^{-1}$. The main source of errors for this region is the contamination by the OH night-sky lines that becomes important given the faintness of the emission.

 There are regions where some splitting of the velocity profiles
 appear. One corresponds to Bubble A where
 at least two velocity components appear (--215  $ \pm 10$ km s$^{-1}$  and --267  $ \pm 10$   km s$^{-1}$) indicating a possible local
 expansion of more than 26 km s$^{-1}$. The other one is the region located where
the NW lobe closes. In this region, there is  splitting in the velocity profiles
 that can be fitted by two velocity components at --233 $ \pm 10$ km s$^{-1}$ and --194  $ \pm 10$ km s$^{-1}$.
 Unfortunately, most of the regions interior to the NW lobe are so faint that
it was impossible to get any reliable profile decomposition. The values of the
 velocity components found suggest an expansion motion of the end of the lobe.
The component at --226 km s$^{-1}$ can be interpreted as the component of the HII region
 at rest (at the systemic velocity of the galaxy) while the component at --194 km s$^{-1}$ could be
 due to the blueshifted region of the lobe. In that case, the expansion velocity
 at the very end of the NW lobe would be: V$_{blue}$ -- V$_{sys}$ = 39 km s$^{-1}$ without any correction for a possible inclination of the bipole axis. This value is smaller than the 75 km s$^{-1}$ value reported by  Afanasiev et al. (2000) for the
expansion velocity of the NW lobe,
 but  it still implies a kinematic age of the nebula  (2 $\times$ 10$^{6}$ yr) of the same order of magnitude.

The [SII] velocity profiles are faint and, consequently, no reliable velocity profile decomposition could be done. Nevertheless, at the end of the NW lobe a single profile shows a heliocentric
 velocity of --190 km s$^{-1}$ confirming the blueshifted value found in the  $H\alpha$
velocity profile of this region.

Figure 10 shows a close-up of the
  map at V$_{HEL}$= --233 km s$^{-1}$ in [SII] where the features discussed above
 are better appreciated. Superimposed on this figure we have marked the boxes
over which we have extracted the velocity profiles and we  also show the obtained radial velocity field at H$\alpha$.
We
 have also constructed integrated position-velocity diagrams from our
  $H\alpha$ and [SII] data cubes. Figure 11  shows the $H\alpha$  position-velocity diagrams and the areas over which they were integrated, both for the S3 complex and a region of IC 1613 where no detectable nebular emission is present.

 In the first case,
 we obtain the position-velocity diagram of the S3 complex and in the second
 case we obtain the position-velocity diagram of the night-sky lines. A
 subtraction of them results in a position-velocity diagram free of night-sky
 contamination. The results of this subtraction are shown in Figure 12. 

As we
 can see from this figure, the velocity width is
 higher in the waist (200 km s$^{-1}$) than in the lobes (150 km s$^{-1}$ for the NW lobe and
 100 km s$^{-1}$ for the SE one). This contradicts the results of Afanasiev et al. (2000) that
 find similar values for the waist but widths of up to 330 and 390 km s$^{-1}$ for the SE
 and NW lobes respectively, i.e., larger velocity widths for the external
 regions. This contradiction could be due to a higher sensibility of the
 observations of Afanasiev et al. (2000) that allows to detect the tenuous fast moving
 regions of the lobes while we are only detecting the brightest and slowest
 regions. However, it would be interesting to examine their data by means of
 position-velocity diagrams similar to the ones obtained in this work.
 It is interesting to note that Afanasiev and collaborators reported an
 expansion velocity of the NW lobe of 75 km s$^{-1}$ whereas their reported velocity widths in 
this region reach 400 km s$^{-1}$ implying induced shock velocities of 200 km s$^{-1}$.
Unless there is an inclination correction (inclination of the axis of the bipolar structure relative to the line of sight of about 70 degrees) it is
 difficult to reconcile the differences between the expansion velocity and the
velocity widths quoted by those authors.

Is the nebular complex S3 a bipolar nebula or a blister? In the first case, the lobes would be shaped by an internal mechanism that favors asymmetric ejections and that focuses the ejected gas. In the blister case, the shape would be due to inhomogeneities in the ISM density. For example, one of the `lobes' would be larger because it could be formed by the stellar winds propagating in a region of decreasing ambient density.
The fact that the
brightest region is located in a place where there is an HI density gradient, as Afanasiev and collaborators have suggested, 
argues in favor of the blister possibility. 
Furthermore, the geometrical models  of bipolar structures, developed in Arias et al. (2000), show that even if we see an inclined bipolar nebula, the dimensions of the lobes remain the same, so that the more plausible explanation for the different dimensions of the lobes is that the NW lobe corresponds to the propagation of a shock in a less dense medium. This scenario could be verified by determining the electron densities of both lobes. We have tried to do so by means of the surface brightness measurements reported in Table 1, but our results are quite uncertain because the NW lobe emission is not easily disentangled from the emission of the superbubble R17.

\section{Conclusions}

- We study the kinematics of the SNR S8, in the galaxy IC 1613, taking into account its environment rich in superbubbles. The SNR S8 appears located at the boundary of the superbubble R10 (according to Paper II's notation) and is seen in projection within superbubble R9. Between the superbubbles R9 and R4, a ridge of obscured material is detected suggesting the presence of dust in
the vicinity of this SNR.

- We discuss the optical appearance of this SNR that does not show the classical filamentary shell characteristic of SNRs; instead, in the visible, radio and X-rays, it shows several of the characteristics of a plerionic-type SNR. 
We give several arguments against this possibility and suggest that we are seeing only a part of a shell-type SNR.

- We obtain the radial velocity field of the optical counterpart of this SNR. We find that the velocity profiles are complex, revealing the presence of several velocity components. The brighter component is at the velocity of neighboring HII regions and has FWHM of about 66  $ \pm 10$ km s$^{-1}$. A broader (FWHM between 66  $ \pm 15$ to 132  $ \pm 15$ km s$^{-1}$) blueshifted velocity component is also identified. However, the velocity widths we find are about half of the values reported previously by Lozinskaya et al. (1998). Also, we do not find the velocity gradient that Lozinskaya and collaborators report for this component.
Our velocity profiles also show  the presence of high velocity gas of low density. In agreement with Lozinskaya et al. (1998) we do not find a clear sign of an expansion pattern.

- Assuming that the optical emission comes from the interaction of a primary blast wave with a dense clump of large dimensions, we were able to derive the  velocity of the shock induced in the clump from the FWZIs of the velocity profiles. We find a value of 170 km s$^{-1}$ for this shock velocity and we estimated the preshock density, the energy released by the supernova and the kinematic age of this SNR. We explore the possibility mentioned before that this SNR could be only a part of a large
diameter shell-type SNR hidden by dust and we find that the energetics is compatible with this hypothesis. However, the possibility that this SNR has
a weird appearance due to the interaction of SN ejecta with a pre-existing cavity cannot be ruled out with the existing data.

- We find that the S3 complex (formed by the bright HII region catalogued by
 Sandage 1971 and several filaments) is constituted of a bright central region
 hosting in its center the WO star and two lobes oriented in the NW-SE direction,
 being the NW lobe longer than the SE lobe.

- We find from the [SII] FP observations that the NW lobe is closed and 
has a clover leaf shape; its major axis is 214 pc long, while the SE lobe is more
 filamentary and has a major axis of 122 pc.

- We have also found that the superbubble R17 in Paper II is seen in
 projection towards the NW lobe, while the superbubble R16 is located near the
end of the NW lobe. These superbubbles have in their interiors and at their boundaries several stellar associations (Borissova et al. 2001) and it is probable these superbubbles are formed by the winds of the massive members of the stellar associations in spite of the close presence of the WO star. On the other hand, the WO star could be ionizing the superbubbles and thus, making them detectable.

- We study the kinematics of this bipolar nebula finding some evidence of
 a possible expansion motion of the NW lobe at 40 km s$^{-1}$, if an inclination
 correction is not important. This value is 50 $\%$ lower than the one reported by
 Afanasiev et al. (2000) but, in any case, neither of those values imply a kinematic
 age in agreement with the duration of the WR phenomena (a few 10$^{5}$ years), 
and even less with the
much shorter 
 duration of the WO phase. However, the privileged position of the WO star is
 the main argument in favor that its progenitor has formed this peculiar
 nebula.

- Our position-velocity diagrams of the S3 system show that the velocity widths
 are larger in the central waist than in the lobes. This contradiction with
 Afanasiev's et al. (2000) results could be due to a probable lower sensitivity of our
 observations.

At last, even if these nebulae (S8 and S3) were formed by only one star in each case, the influence of those stars upon the ISM of IC 1613 is quite remarkable.

\acknowledgments

The authors thank the Bulgarian Academy of Sciences and the Instituto de Astronom\'ia, UNAM, for their support in the developing of this collaboration.
MR thanks also Mrs. J. Benda for reading the manuscript.

      Part of this work was supported by  grants 27984-E of CONACYT and IN122298 of DGAPA-UNAM.

\clearpage

\figcaption{ H$\alpha$ velocity map at  the 
heliocentric velocity of --253 km s$^{-1}$ of the central region of the galaxy IC 1613. 
The locations of the selected nebulae studied here are clearly marked.
The orientation and the scale are indicated up and down to the right,
respectively.
 \label{fig1}}

\figcaption{ Idem but for the [SII]($\lambda$ 6731 \AA) line. Note that the H$\alpha$ velocity 
map has a better spatial resolution (1$\arcsec$.18 px$^{-1}$) compared with 
the [SII] one (2$\arcsec$.36 pix$^{-1}$). 
\label{fig2}}

\figcaption{Close-up of the SNR S8 H$\alpha$ emission. The isophotes show the elongated form of the SNR. Part of the HII region S7 is seen to the north. The orientation and the scale are indicated. 
\label{fig3}}

\figcaption{ Typical radial velocity profile of the SNR S8 at H$\alpha$ (upper case) and  at [SII]($\lambda$6731 \AA) (lower case). For the [SII] profile, the secondary peak corresponds to the [SII] ($\lambda$6717 \AA) line. 
\label{fig4}}

\figcaption{ Radial velocity field of the SNR S8. The square boxes correspond to the regions over which the integration was done in order to extract the velocity profiles. Only the main components of the profile fitting are shown.
\label{fig5}}

\figcaption{Close-up of the field around the nebula S3 showing the position of the stellar associations found in Borissova et al. (2001) superimposed
onto the limits of the superbubbles R12, R16 and R17. Likewise the spatial position 
of the WO star is indicated. The orientation and the scale are also indicated.
\label{fig6}}

\figcaption{H$\alpha$ radial velocity maps of the S3 complex (high contrast). The numbers appearing in the lower left corner of each map correspond to the heliocentric velocities. 
\label{fig7}} 

\figcaption{ H$\alpha$ radial velocity maps of the S3 complex (low contrast).
The superbubbles R16 and R17 discussed in the text, are visible in the velocity maps at  V$_{HEL}$=--272, --253 and --234 km s$^{-1}$.
\label{fig8}}

\figcaption{ [SII] radial velocity maps of the S3 complex. The arrow in the map at V$_{HEL}$=--233 km s$^{-1}$ shows where the NW lobe closes.
\label{fig9}}

\figcaption{Radial velocity field of S3 superimposed on the [SII] image map at V$_{HEL}$=--233 km s$^{-1}$. The square boxes show the integration areas over which the velocity profiles were extracted. The dashed lines show the appearance of the bipolar nebula. The orientation and scale are indicated.
\label{fig10}}

\figcaption{Position-velocity diagrams of the bipolar nebula S3 (left) and of an `empty' region that gives us the night-sky contaminating emission (right). The images shown in the upper part of this figure show the regions where the position-velocity diagrams were obtained by integrating over their longer axis. 
\label{fig11}}

\figcaption{Nebular position-velocity diagrams free from night-sky emission, obtained from the subtraction of the position-velocity diagrams shown in Fig. 12. In the upper panel we show the direct image of the bipolar nebula at the same scale.
\label{fig12}}

\clearpage

\begin{deluxetable}{cc}
\scriptsize
\tablewidth{0pc}
\tablenum{1}
\tablecaption{Surface brightness values of the different regions of S3.}
\tablehead{Region &  S(H$\alpha$)~ ergs cm$^{-2}$ s$^{-1}$ sr$^{-1}$ }
\startdata
NW LOBE & (2.80  $ \pm 0.61$ ) $\times$ 10$^{-6}$      \\
SE LOBE  &  (8.11  $ \pm 1.95$ )$\times$ 10$^{-6}$   \\
WAIST   & (4.45  $ \pm 1.0 $) $\times$ 10$^{-5}$   \\
\enddata
\end{deluxetable}
\clearpage

\end{document}